\documentclass[prb,twocolumn,secnumarabic,showpacs,amsmath,amssymb,floatfix,superscriptaddress,longbibliography,aps]{revtex4-1}

\usepackage{amssymb}
\usepackage{times}
\usepackage{graphicx}
\usepackage{epsfig}
\usepackage{color,soul}
\usepackage{multirow}
\usepackage{threeparttable}
\usepackage{bm}
\usepackage{epstopdf}
\usepackage{changes}
\usepackage{lineno}

\begin{document}
\title{Lattice dynamics and coupled quadrupole-phonon excitations in CeAuAl$_3$}
\author{B.-Q. Liu}
\email{liuqiong414@163.com}
\affiliation{Key Laboratory of Neutron Physics, Institute of Nuclear Physics and Chemistry, CAEP, Mianyang 621900, PR China}
\affiliation{J{\"u}lich Centre for Neutron Science (JCNS) at Heinz Maier-Leibnitz Zentrum (MLZ), Forschungszentrum J{\"u}lich GmbH, Lichtenbergstra{\ss}e 1, 85748 Garching, Germany}
\author{P. \v{C}erm\'{a}k}
\affiliation{J{\"u}lich Centre for Neutron Science (JCNS) at Heinz Maier-Leibnitz Zentrum (MLZ), Forschungszentrum J{\"u}lich GmbH, Lichtenbergstra{\ss}e 1, 85748 Garching, Germany}
\affiliation{Charles University, Faculty of Mathematics and Physics, Department of Condensed Matter Physics, Ke Karlovu 5, 121 16, Praha, Czech Republic}
\author{C. Franz}
\affiliation{Physik-Department, Technische Universit\"{a}t M\"{u}nchen, 85748 Garching, Germany}
\affiliation{Heinz Maier-Leibnitz Zentrum (MLZ), Technische Universit\"{a}t M\"{u}nchen, 85748 Garching, Germany}
\author{C. Pfleiderer}
\affiliation{Physik-Department, Technische Universit\"{a}t M\"{u}nchen, 85748 Garching, Germany}
\author{A. Schneidewind}
\affiliation{J{\"u}lich Centre for Neutron Science (JCNS) at Heinz Maier-Leibnitz Zentrum (MLZ), Forschungszentrum J{\"u}lich GmbH, Lichtenbergstra{\ss}e 1, 85748 Garching, Germany}

\begin{abstract}
We report first principles calculations of the structural parameters and phonon dispersion of the tetragonal non-centrosymmetric heavy fermion compound CeAuAl$_3$. Taking into account weak magnetoelastic interactions of the rare-earth (RE) ions with the spectrum of phonons, we obtain an analytical expression for the hybridization of quadrupole excitations and phonons from the poles of the one-phonon Green-function. In the paramagnetic phase, we predict the formation of mixed modes that may be observed by inelastic neutron scattering. Our results show that magnetoelastic interactions, albeit being moderate, play an important role in CeAuAl$_3$. This suggests that magnetoelastic interactions may be equally important in a wide range of related compounds.
\end{abstract}

\maketitle

\section{Introduction}
Cerium-based heavy fermion compounds with the general formula Ce$TX_3$, where $T$ is a transition metal element and $X=$ Si, Ge, or Al feature a broad variety of different ground states, such as heavy-fermion behavior, pressure-induced superconductivity, or magnetic properties resulting from the interplay between the Kondo effect and RKKY interactions.\cite{Y.Aoki2000,X.D.Hu2009,A.D.Hillier2012,V.K.Anand2011,M.Smidman2013,M.Klicpera2015,M.Klicpera2014,N.Kimura2005,I.Sugitani2006,R.Settai2007,S.Mock1999}
Poised at the border of a structural instability, these systems are often reported to grow in several related structure types such as CeCuAl$_3$, \cite{Bauer1987,Mentink1993,M.Klicpera2014}
exhibiting also structural transitions \cite{M.Klicpera2014}.
Moreover, crystal electric field (CEF) effects play an essential role for the physical properties of these strongly correlated electrons systems. In particular the coupling of the CEFs to homogeneous and inhomogeneous distortions of the lattice represented by phonons appears to be of primary importance as reported in CeCuAl$_3$ and CeAuAl$_3$.\cite{D.T.Adroja2012,Y.Aoki2000}

Typically such magnetoelastic interactions are neglected, assuming that they influence the system only weakly. However, in case of strong magneto-elastic coupling, phenomena such as mixed-mode excitations of phonons and quadrupolar excitations have been reported when the dispersion curves intersect in momentum space. This may result in an anticrossing that may be observed directly in inelastic neutron scattering (INS). Several studies have reported putative evidence of an anticrossing of acoustic phonons and quadrupole excitations, e.g., in PrAlO$_3$ \cite{R.J.Birgeneau1974}, TbVO$_4$ \cite{Hutchings1975}, TmVO$_4$ \cite{Kjems1975}. In contrast, in intermetallic compounds an anticrossing could only be inferred indirectly as reported for PrNi$_5$ \cite{V.L.Aksenov1983}.

When the magnetoelastic coupling becomes very strong, a completely different excitation spectrum may be expected resulting in the formation of a bound state between low-lying optical phonons and a crystal field excitation.\cite{P.Fulde1985} Such bound states have been reported in CeAl$_2$ \cite{M.Loewenhaupt1979}, CeCu$_2$ \cite{M.Loewenhaupt2003}, and CePd$_2$Al$_2$ \cite{L.C.Chapon2006}, as interpreted by a theoretical model proposed by Thalmeier and Fulde.\cite{P.Thalmeier1982,P.Thalmeier1984} In a related study, Adroja \textit{et al.} \cite{D.T.Adroja2012} recently reported the observation of a bound state in CeCuAl$_3$, indicating the presence of particularly large magnetoelastic coupling. Yet, the same anomalous property, using the same neutron time-of-flight spectroscopic technique, could not be observed in the isostructural sibling CeAuAl$_3$,\cite{Adroja2015} which displays a Kondo temperature $T_K=3.5\,{\rm K}$ and an antiferromagnetic transition at $T_N=1.32\,{\rm K}$.

In the study reported here, we present first principles calculations of the structural properties and phonon dispersion of CeAuAl$_3$. Taking into account weak magnetoelastic coupling, we predict substantial hybridization of quadrupole excitations with phonons amenable to observation by inelastic neutron scattering.\cite{P.Cermak}

\section{Structural properties and lattice dynamics}
First principles calculations have been carried out by means of the Vienna ab initio simulation package (VASP) \cite{G.Kresse1993,G.Kresse1996}, using the frozen-core projector augmented wave (PAW) method. GGA descriptions for the exchange-correlation functional were used with a cutoff energy of 500 eV in a plane-wave basis expansion. The Brillouin zone (BZ) was sampled by 14$\times$14$\times$7 mesh of $k$-points, as determined according to the Monkhorst-Pack scheme. The tetrahedron method with Bl\"ochl corrections was employed for the energy calculation, and Methfessel-Paxton's Fermi-level smearing was used to accelerate electronic structure relaxation. A quasi-Newton algorithm was adopted for the geometric relaxations, with a convergence criterion of the Hellmann-Feynman force of 0.01 eV/{\AA}.

\begin{figure}
\centering
\includegraphics[width=8.5cm]{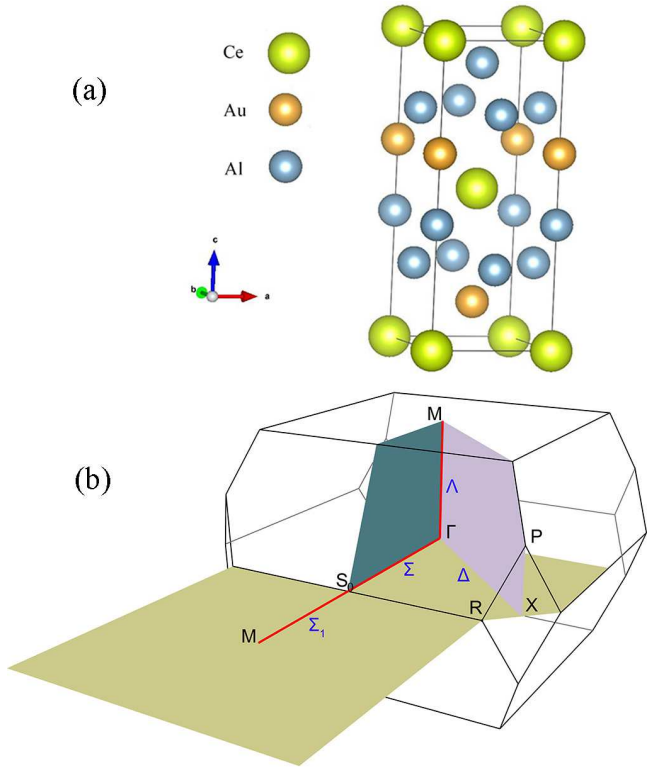}
\hspace{0.5cm}
\caption{(Color online) (a) Crystal structure of CeAuAl$_3$. Ce, Au and Al atoms are shown in yellow, orange, and blue shading, respectively. (b) Brillouin zone of the body-centered tetragonal lattice of CeAuAl$_3$ with $c>a$.}
\label{figure-1}
\end{figure}

\begin{figure}
\centering
\includegraphics[width=8.5cm]{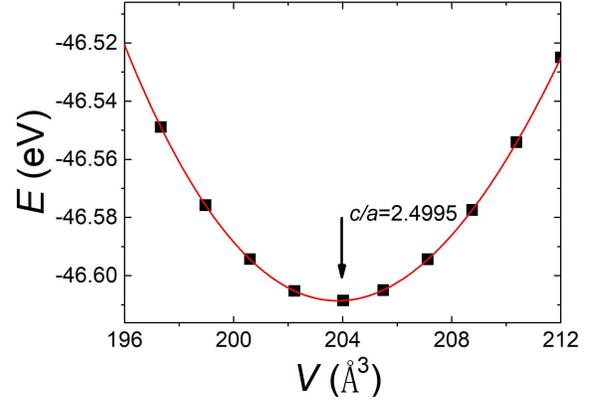}
\hspace{0.5cm}
\caption{(Color online) Total energy of CeAuAl$_3$ as a function of unit cell volume. The solid curve represents a least-squares fit to the third-order Birch-Murnaghan equation of state, Eq.(1).}
\label{figure-2}
\end{figure}

\begin{figure}
\centering
\includegraphics[width=8.5cm]{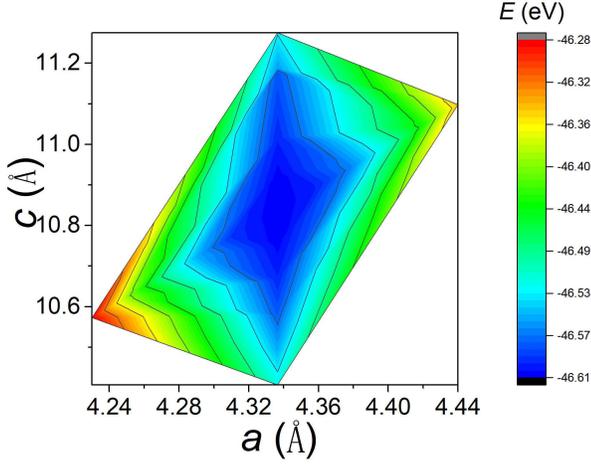}
\hspace{0.5cm}
\caption{(Color online) Depiction of the total energy of CeAuAl$_3$ as a function of the lattice parameters $a$ and $c$. Lattice parameters at the minimum are in excellent agreement with experiment; for values see Table \ref{table-1}. }
\label{figure-3}
\end{figure}

CeAuAl$_3$ crystallizes in the non-centrosymmetric BaNiSn$_3$-type tetragonal structure (space group is $I4mm$, No.107). The conventional unit cell contains 10 atoms, as shown in Figure \ref{figure-1}(a). The lattice parameters are $a_0=4.3364\,{\rm \AA}$, and $c_0=10.85\,{\rm \AA}$, corresponding to a $c/a$ ratio of 2.5021.\cite{C.Franz} For geometry optimizations we started with the experimental geometries and calculated the dependence of the total energy ($E$) on the volume ($V$). For the tetragonal lattice the calculations started with a fixed lattice parameter $a=4.3364\,{\rm \AA}$ and the volume was adjusted by choosing different $c/a$ ratios. Using a least-squares fit of the $E$-$V$ curve (shown in Fig.\ref{figure-2}) to the third-order Birch-Murnaghan equation of state (EOS),\cite{F.Birch1947}
\begin{eqnarray}
E(V)&=&E_0+\frac{9V_0B_0}{16}\Big\{\Big[(V_0/V)^{\frac{2}{3}}-1\Big]^3B'_0 \nonumber\\
&+&\Big[(V_0/V)^{\frac{2}{3}}-1\Big]^2\Big[6-4(V_0/V)^{\frac{2}{3}}\Big]\Big\},
\end{eqnarray}
the equilibrium energy $E_0$, the bulk modulus $B_0$ and its derivative $B'_0$ at $P=0$ and $T=0$ were determined. The value of $c/a$ at the energy minimum is denoted as $\xi$. Similarly, we fixed the $c/a$ ratio as $\xi$=2.4995 and adjusted the volume by changing the lattice parameter $a$. Figure \ref{figure-3} shows the dependence of the total energy on lattice parameters. After the second EOS fit, the optimized lattice parameters were obtained as $a$=4.335 \AA, $c$=10.844 \AA, with $c/a=$ 2.5012, and an equilibrium atomic volume $V_0=203.81\,{\rm \AA}^3$, as listed in Table \ref{table-1}. Comparison with previous experimental data, also shown in Table \ref{table-1}, reveals excellent agreement.\cite{C.Franz}

\begin{table}[htbp]
\caption{Comparison of the lattice parameters, $c/a$ ratio and unit cell volume of CeAuAl$_3$ observed experimentally and reported in Refs.\,\cite{Adroja2015,C.Franz} with ab-initio values calculated as part of our study.}
\begin{tabular}{lcccc}
\hline
                                   &     $a$ (\AA) &  $c$ (\AA) &  $c/a$     &   $V_0$(\AA$^3$)\\
\hline
Expt. at 300 mK \cite{Adroja2015}  &     4.3105    & 10.7965    &  2.5047    &   200.60         \\
Expt. at 9 K \cite{Adroja2015}     &     4.3172    & 10.8090    &  2.5037    &   201.46          \\
Expt. at 300 K \cite{C.Franz}                &     4.3364    & 10.85      &  2.5021    &   204.03           \\
\hline
Our DFT study                      &     4.3354    & 10.8436    &  2.5012    &   203.81            \\
\hline
\end{tabular}
\label{table-1}
\end{table}

\begin{figure}
\centering
\includegraphics[width=8.5cm]{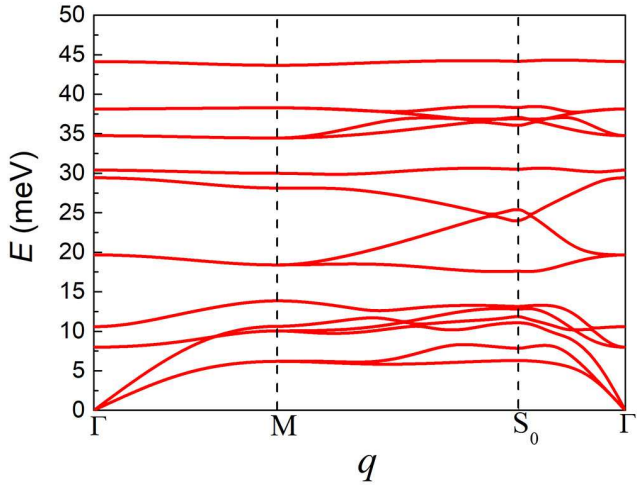}
\hspace{0.5cm}
\caption{Phonon dispersion for CeAuAl$_3$ along high symmetry lines $\Gamma-$M$-$S$_0-\Gamma$, where $\Gamma=(0,0,0)$, M$=(1/2,1/2,-1/2)$, and S$_0=(-\eta,\eta,\eta)$ in primitive basis, with $\eta=(1+a^2/c^2)/4$.}
\label{figure-4}
\end{figure}

Phonon calculations for CeAuAl$_3$ were performed using the finite difference method. A $2\times2\times2$ supercell was constructed and the code of Phonopy was used.\cite{Togo2008,Togo2010}  For CeAuAl$_3$, there are five atoms in a primitive cell and the full phonon dispersion of CeAuAl$_3$ consists of 15 branches comprising three acoustic and twelve optical modes. However, at certain high symmetry points and along certain high symmetry directions, such as $\Gamma-$M, mode degeneracies may be seen. Results of our calculations along the high-symmetry direction $\Gamma-$M$-$S$_0-\Gamma$ in the tetragonal Brillouin zone are shown in Fig.\,\ref{figure-4}.

Using group theoretical techniques, the irreducible representations for the high symmetry points $\Gamma$, M, and S$_0$ may be obtained, given by
\begin{eqnarray}
\Gamma &=&4\Gamma_1+\Gamma_3+5\Gamma_5, \nonumber\\
M&=&4\Gamma_1+\Gamma_3+5\Gamma_5,\nonumber\\
S_0&=&9A^\prime+6A^{\prime\prime}.\nonumber
\end{eqnarray}
Since all the representations of $C_{4v}$ are one dimensional except for $\Gamma_5$, we find that $\Gamma_1$ and $\Gamma_3$ are single modes without degeneracy, while the two transverse acoustic branches along $\Gamma-$M are degenerate.

\section{Coupled quadrupole-phonon excitations}
The magnetoelastic properties of rare earth systems originate mainly in the interaction of 4$f$-electrons with total angular momentum $\bm{J}$ with the crystal electric field. As mentioned above, a lot of effort has been devoted to the study of coupled quadrupole-phonon excitations reflecting magnetoelastic interactions.\cite{R.J.Birgeneau1974,Hutchings1975,Kjems1975,V.L.Aksenov1983} Several theoretical derivations for the mixed-mode excitation dispersion have been reported. Using a diagrammatic method the mixed-mode excitation dispersion may be inferred from the poles of the dynamical susceptibility of the system of RE ions.\cite{P.Thalmeier1991,P.Thalmeier1975}  Another important approach was proposed by Aksenov \textit{et al.} \cite{V.L.Aksenov1981}, using the equation of motion method for the Green-function (GF) as a basis to describe coupled quadrupole-phonon excitations. In our study both methods give the same results, since some secondary effects are not included as described below. We will treat $f$-electrons as strictly localized, because first CEF excitation at $E$=59.2\,K \cite{Adroja2015} lies well above $T_\textrm{K}$.

The Hamiltonian for rare-earth compounds, taking into account magnetoelastic interactions, may be written as
\begin{eqnarray}
H=H_{\rm Lattice}+\sum_{f=1}^N{H_{\rm CEF}^f},\\
H_{\rm CEF}^f=H_{\rm CEF}^0+H_{\rm me}^I,
\label{hamil}
\end{eqnarray}
where $H_{\rm Lattice}$ describes phonons in the harmonic approximation, $H_{\rm CEF}^f$ denotes the interaction of the spin $\bm{J}$ of the $f$-electron containing atom with the crystal field due to the surrounding atoms, and the summation over $f$ extends over all atoms in the crystal. The term $H_{\rm CEF}^0$ is the conventional crystal field Hamiltonian for the stationary lattice, which may be expressed in terms of Steven's operators. $H_{me}^I$ is the single-ion magnetoelastic interaction, coupling the spin system to the strain. This simple model allows to describe the elementary excitations, which may be observed by inelastic neutron scattering where the differential neutron scattering cross section may be expressed in terms of the phonon Green-function \cite{W.Marshall1971}.

Although there is no inversion symmetry (I) at the Ce-site, the usual CEF potential for I-symmetric cites may be used in CeAuAl$_3$ as an approximation. The CEF Hamiltonian is then given by \cite{Adroja2015}
\begin{equation}
H_{\rm CEF}^0=B_2^0O_2^0+B_4^0O_4^0+B_4^4O_4^4,
\end{equation}
where $B_n^m$ and $O_n^m$ are the CEF parameters and Steven's operators, respectively.\cite{Balcar1989,Hutchings1964} The sixfold degenerate Ce$^{3+}$ ($J=\frac{5}{2}$) states, 4$f^1$, splits into three doublets in the paramagnetic phase. The CEF parameters of CeAuAl$_3$ have been determinated by inelastic neutron scattering.\cite{Adroja2015} They are given by $B_2^0=1.2208(130)$ meV, $B_4^0=-0.0021(3)$ meV, and $B_4^4=0.2555(2)$ meV. With this set of CEF parameters, one obtains the eigenvalues $E_{\Gamma_6}=-10.0184$ meV, $E_{\Gamma_7^1}=-4.84099$ meV, and $E_{\Gamma_7^2}=14.8594$ meV, and the corresponding eigenvectors
\begin{eqnarray}
|\Gamma_6\rangle &=& |\pm\frac{1}{2}\rangle,\nonumber\\
|\Gamma_7^1\rangle &=& -\alpha|\mp\frac{3}{2}\rangle+\beta|\pm\frac{5}{2}\rangle,\nonumber\\
|\Gamma_7^2\rangle &=& \alpha|\mp\frac{5}{2}\rangle+\beta|\pm\frac{3}{2}\rangle,
\end{eqnarray}
with $\alpha=0.927$, and $\beta=0.375$. At low temperatures, only the lowest-lying level is occupied. The excitation energy for the $|\Gamma_6\rangle\rightarrow|\Gamma_7^1\rangle$ transition corresponds to $E_{\Gamma_6\Gamma_7^1}=5.1$ meV, which is a factor of five smaller as compared to the energy of the $|\Gamma_6\rangle\rightarrow|\Gamma_7^2\rangle$ transition. This suggests roughly cubic CEFs ($B_4^0$ is very small) with the doublet $|\Gamma_6\rangle$ forming a quasi-quartet, consistent with the cage-like environment of Ce in Fig. 1(a). In the following we consider the transition $|\Gamma_6\rangle\to|\Gamma_7^1\rangle$ between the two lowest-lying states only which crosses the acoustic phonons.

The single-ion terms in the magnetoelastic Hamiltonian account for the direct coupling between the deformations of the lattice and the 4$f$ shell. This Hamiltonian may be constructed according to group theory.\cite{E.Callen1963,E.Callen1965} 
For the spin functions $\mathcal{S}_i^{\Gamma,j} (i=1,2,\cdots,n)$ which form a basis for the $n$-dimensional representation $\Gamma$, where different sets are denoted by $j=$1, 2, as shown in Table II, the single-ion contributions $H_{\rm me}^I$ to the magnetoelastic Hamiltonian becomes
\begin{equation}
H_{\rm me}^I=-\sum_{\Gamma}{\sum_{j,j'}{\widetilde{B}_{jj'}^\Gamma(f)\sum_i{\epsilon_i^{\Gamma,j}\mathcal{S}_i^{\Gamma,j'}(f)}}},
\end{equation}
where $\widetilde{B}_{jj'}^\Gamma$ represents a material-specific phenomenological magnetoelastic coupling constant, and the number of single-ion magnetoelastic coupling constants for the tetragonal point group is listed in the last two columns of Table III. $\epsilon_i^{\Gamma,j}$ are linear combinations of the first-order strain components $\epsilon_{\mu\nu}$ ($\mu,\nu=x,y,z$). The shear strain is defined as $\epsilon_{\mu\nu}=\frac{1}{2}(\frac{\partial u_\nu}{\partial\mu}+\frac{\partial u_\mu}{\partial\nu})$. As a constraint we assume that all antisymmetric strains such as $\frac{1}{2}(\frac{\partial u_y}{\partial x}-\frac{\partial u_x}{\partial y})$, which correspond to homogeneous rotations of the crystal, vanish. This assumption is reasonable, because according to Fulde \cite{Fulde1979}, although in zero external field the rotational magnetoelastic interaction contributes to the phonon dispersion curves for finite wave vectors, it only gives corrections of a few percent. An additional justification presented in Ref. \onlinecite{P.Thalmeier1975} concerns that rotational effects cannot be included correctly, if the full crystal level scheme is not taken into account. As Eq.(6) has been derived using homogeneous strains, it is only valid for long-wavelength acoustic phonons.\cite{P.Thalmeier1991}

\begin{table*}[htbp]
\caption{Strain functions and single-ion spin operators for the tetragonal system.}
\begin{tabular}{ll}
\hline
Strain functions $\epsilon_i^{\Gamma,j}$&   Single-ion operators $\mathcal{S}^{\Gamma,j'}(f)$\\
\hline
$\epsilon^{\alpha 1}\equiv\frac{\sqrt{3}}{3}(\epsilon_{xx}+\epsilon_{yy}+\epsilon_{zz})$ &$1$\\
$\epsilon^{\alpha 2}=\sqrt{\frac{2}{3}}[\epsilon_{zz}-\frac{1}{2}(\epsilon_{xx}+\epsilon_{yy})]$ &$3J_z^2-J(J+1)$\\
$\epsilon^\gamma=\frac{\sqrt{2}}{2}[\epsilon_{xx}-\epsilon_{yy}]$ & $J_x^2-J_y^2=\frac{1}{2}(J_+^2+J_-^2)$\\
$\epsilon^\delta=\sqrt{2}\epsilon_{xy}$ &  $P_{xy}=\frac{1}{2}(J_xJ_y+J_yJ_x)$\\
$\epsilon_1^\epsilon=\sqrt{2}\epsilon_{yz}$ &  $P_{yz}=\frac{1}{2}(J_yJ_z+J_zJ_y)$\\
$\epsilon_2^\epsilon=\sqrt{2}\epsilon_{zx}$ &  $P_{zx}=\frac{1}{2}(J_xJ_z+J_zJ_x)$
\\
\hline
\end{tabular}
\end{table*}

According to Table II and Table III, the single-ion magnetoelastic Hamiltonian for the tetragonal symmetry is given by
\begin{eqnarray}
H_{\rm me}^I(f)=&-&\widetilde{B}^{\alpha_1}\epsilon^{\alpha 1}\Big[3J_z^2-J(J+1)\Big]\nonumber\\
&-&\widetilde{B}^{\alpha_2}\epsilon^{\alpha 2}\Big[3J_z^2-J(J+1)\Big]\nonumber\\
&-&\widetilde{B}^\gamma\frac{\sqrt{2}}{2}(\epsilon_{xx}-\epsilon_{yy})(J_x^2-J_y^2)\nonumber\\
&-&\widetilde{B}^\delta\frac{\sqrt{2}}{2}\epsilon_{xy}(J_xJ_y+J_yJ_x)\nonumber\\
&-&\widetilde{B}^\epsilon\frac{\sqrt{2}}{2}\epsilon_{yz}(J_yJ_z+J_zJ_y)\nonumber\\
&-&\widetilde{B}^\epsilon\frac{\sqrt{2}}{2}\epsilon_{xz}(J_xJ_z+J_zJ_x).
\end{eqnarray}

\begin{table*}[htbp]
\begin{threeparttable}
\caption{Group table for crystal point group. Pairs of functions in a square bracket form a two-dimensional irreducible representation. $N_{el}$ denotes the number of elastic constants, and $N_{me}$ denotes the number of single-ion magnetoelastic coupling constants.}
\begin{tabular}{cccccccc}
\toprule
System &Point group & Basis functions &Dimensionality &\multicolumn{2}{p{50pt}}{$N_{el}$} &\multicolumn{2}{p{50pt}}{$N_{me}$}\\
\hline
\multirow{4}{*}{Tetragonal} &
\multirow{4}{*}{$4mm$}     &
$x^2+y^2+z^2$, $(\sqrt{3}/2)(z^2-\frac{1}{3}r^2)$ &1 &3  &\multirow{4}{*}{6}    &3   &\multirow{4}{*}{6}  \\
& &$\frac{1}{2}(x^2-y^2)$                         &1 &1  &                      &1   &  \\
& &$xy$                                           &1 &1  &                      &1   &  \\
& &$[yz,xz]$                                      &2 &1  &                      &1   &  \\
\toprule\\
\end{tabular}
\end{threeparttable}
\end{table*}
Since the single-ion mangetoelastic Hamiltoninan is sufficiently general when considering the main effects,\cite{P.Morin} our study avoids additional complications caused by two-ion magnetoelastic interactions which may lead to structural and magnetic phase transitions.\cite{Dohm1975}

In conventional linear strain theory, i.e., neglecting the second order magnetoelastic interaction as well as the linear rotational interaction,\cite{Dohm1975,V.L.Aksenov1981} and expanding the crystal field potential in powers of the lattice deformations, the Hamiltonian given in Eq.\,\ref{hamil} may be written as
\begin{equation}
H_{\rm CEF}^f=H_{\rm CEF}^0+\sum_\mathbf{q}{(a_\mathbf{q}+a_{-\mathbf{q}}^\dagger)V(\bm{J},\bm{q})\exp{(i\bm{q}\cdot\bm{R})}},
\end{equation}
where $a_\mathbf{q}^\dagger$ and $a_\mathbf{q}$ are creation and annihilation operators for the phonons with wave vector $\bm{q}$. Usually the experimental data, even in the presence of an applied magnetic field may be fitted well by the term $V(\bm{J},\bm{q})$ alone. The magnetoelastic interaction operator permits the hybridization of quadrupolar excitations and transverse acoustic phonons ($c_{44}$ mode) in several symmetry directions of the reciprocal lattice. For instance:\\
(i) M-direction:
\begin{eqnarray}
V(\bm{J},\bm{q})\sim e_x(q_z)q_z[J_xJ_z+J_zJ_x]+e_y(q_z)q_z[J_yJ_z+J_zJ_y], \nonumber
\end{eqnarray}
(ii) $[1,0,0]$ direction:
\begin{eqnarray}
V(\bm{J},\bm{q})\sim e_z(q_x)q_x[J_xJ_z+J_zJ_x], \nonumber
\end{eqnarray}
(iii) $[0,1,0]$ direction:
\begin{eqnarray}
V(\bm{J},\bm{q})\sim e_z(q_y)q_y[J_yJ_z+J_zJ_y],
\end{eqnarray}
where $\bm{e}(\bm{q})$ is the polarization vector for the phonons. For different wave vectors $[\zeta,0,0]$, $[0,\zeta,0]$ and $[0,0,\zeta]$, $V(\bm{J},\bm{q})$ may be related to the dynamical matrix $\bm{M}$ for the acoustic modes, with the matrix element $M_{ik}=\sum_{j,l}{c_{ijkl}q_jq_l}$. The solutions are listed in Table IV. In addition to the mixing for transverse phonons, the magnetoelastic interaction could also lead to the appearance of mixed modes for longitudinal phonons with $\textbf{q}\cdot\textbf{e}=1$. For example, the quadrupole operator $3J_z^2-J(J+1)$ has non-zero matrix elements for the CEF transitions $|\Gamma_7^{1A(B)}\rangle\rightarrow|\Gamma_7^{2B(A)}\rangle$, and allows the mixing for longitudinal modes at elevated temperatures.

\begin{table}[htbp]
\caption{Long-wavelength acoustic modes generated by shear distortions of the crystal structure representing solutions of the dynamic matrix $\bm{M}$ for three wave vectors $[\zeta,0,0]$, $[0,\zeta,0]$ and $[0,0,\zeta]$ are tabulated, where $\rho$ is the density, $c_{ij}$ are elastic constants. The subscripts on $\omega$ denote the eigenvectors of $\bm{M}$ representing the corresponding motions of the atoms.}
\begin{tabular}{ccc}
\hline
$\bm{q}=[\zeta,0,0]$                 &     $\bm{q}=[0,\zeta,0]$                    &   $\bm{q}=[0,0,\zeta]$                       \\
\hline
$\rho\omega_{[100]}^2=c_{11}\zeta^2$ &     $\rho\omega_{[010]}^2=c_{22}\zeta^2$    &   $\rho\omega_{[001]}^2=c_{33}\zeta^2$         \\
$\rho\omega_{[001]}^2=c_{44}\zeta^2$ &     $\rho\omega_{[001]}^2=c_{44}\zeta^2$    &   $\rho\omega_{[010]}^2=c_{44}\zeta^2$          \\
$\rho\omega_{[010]}^2=c_{66}\zeta^2$ &     $\rho\omega_{[100]}^2=c_{66}\zeta^2$    &   $\rho\omega_{[100]}^2=c_{44}\zeta^2$           \\
\hline
\end{tabular}
\end{table}

Following the strategy developed by Aksenov \textit{et al.},\cite{V.L.Aksenov1981,V.L.Aksenov1983} the differential neutron cross section may be inferred from the one-phonon Green-function,
\begin{equation}
D(\bm{q},\omega)=\Big[\Big(D^0(\bm{q},\omega)\Big)^{-1}-\sum_{mn}{G_{mn}(\bm{q},\omega)}\Big]^{-1},
\end{equation}
where $D^0(\bm{q},\omega)=\frac{2\omega_0}{\omega_\mathbf{q}^2-\omega_0^2}$ is the phonon GF in the harmonic approximation, with $\omega_0$ the frequency of the lattice vibration, and $G_{mn}(\bm{q},\omega)$ in the paramagnetic phase describes single-ion quadrupole excitation
\begin{equation}
G_{mn}(\bm{q},\omega)=\frac{E_{mn}V_{mn}(\mathbf{J},\mathbf{q})V_{nm}(\mathbf{J},\mathbf{q})(f_m-f_n)^2}{\omega_\mathbf{q}^2-E_{mn}^2},
\end{equation}
where $E_{mn}=E_n-E_m$ are the energies of transitions between CEF levels, $f_m=\exp{(-\beta E_m)}/\sum_m{\exp{(-\beta E_m)}}$ is the Boltzmann population factor, and $V_{mn}(\mathbf{J},\mathbf{q})=\langle m|V(\bm{J},\bm{q})|n\rangle$ are the matrix elements of the operator $V(\bm{J},\bm{q})$ (9). For the $\Gamma-$M direction, only the matrix elements determined by $\Gamma_6^A-\Gamma_7^{1B}$ and $\Gamma_6^B-\Gamma_7^{1A}$ differ from zero, and may be written in the form
\begin{equation}
|V_{\Gamma_6^A\Gamma_7^{1B}}|=|V_{\Gamma_6^B\Gamma_7^{1A}}|=2\sqrt{2}\alpha\widetilde{B}^\epsilon|\bm{q}|,
\end{equation}
where the superscripts $A$ and $B$ are used to distinguish the doublets $|\Gamma_6\rangle$ and $|\Gamma_7^1\rangle$.

Using Eqs.\,(10) to (12), the energy of the coupled quadrupole-phonon excitation dispersion, $\omega(\bm{q})$, may be determined by the poles of the Green-Function in Eq.\,(10), i.e., by setting the denominator of Eq.\,(10) equal to zero \cite{Dohm1975}
\begin{eqnarray}
(\omega_\mathbf{q}^2)^2-\omega_\mathbf{q}^2E_{\Gamma_6\Gamma_7^1}^2-\omega_0^2\omega_\mathbf{q}^2+\omega_0^2E_{\Gamma_6\Gamma_7^1}^2\nonumber\\
-\frac{\hbar^2}{M_c}E_{\Gamma_6\Gamma_7^1}(|V_{\Gamma_6^A\Gamma_7^{1B}}|^2+|V_{\Gamma_6^B\Gamma_7^{1A}}|^2)=0.
\end{eqnarray}
One may then obtain the collective excitation spectra for the coupled quadrupole-phonon modes in the paramagnetic phase as follows
\begin{eqnarray}
\omega_{\mathbf{q}\pm}^2&= &\frac{1}{2}(E_{\Gamma_6\Gamma_7^1}^2+\omega_0^2)\mp\Big[\Big(\frac{1}{2}(E_{\Gamma_6\Gamma_7^1}^2-\omega_0^2)\Big)^2\nonumber\\
&+&\frac{\hbar^2}{M_c}E_{\Gamma_6\Gamma_7^1}(|V_{\Gamma_6^A\Gamma_7^{1B}}|^2+|V_{\Gamma_6^B\Gamma_7^{1A}}|^2)\Big]^{\frac{1}{2}},
\end{eqnarray}
where $f_{\Gamma_6}=1$ and $f_{\Gamma_7}=0$, and $M_c$ is the unit cell mass.

For the modeling of the acoustic phonon dispersion a one-dimensional chain is used $\omega_0=K\sqrt{1-\cos(q)}$, where $K$ is related to the amplitude of the branch and determined experimentally. Therefore, using equations (12) and (14), one may infer the magnetoelastic constant from experiment. The calculated dispersion curves of the mixed modes, i.e., the coupled quadrupole-phonon excitations described by Eq.\,(14) are shown in Fig.\,5.

Recently, a series of neutron scattering experiments of single crystal CeAuAl$_3$ have been performed at the triple axis spectrometers PUMA \cite{PUMA} and PANDA \cite{PANDA} at the Maier-Leibnitz Zentrum (MLZ) in Garching, Germany. Here the profond anticrossing as calculated in this paper was observed with an estimated effective magnetoelastic coupling constant $g_{AC}=\frac{(\widetilde{B}^\epsilon)^2}{c_{44}\Omega}$ of 12.1(2) $\mu$eV, where $\Omega$ is the volume of the primitive cell. This provides compelling evidence of magnetoelastic interactions in this compound. These data will be reported in a separate paper \cite{P.Cermak}. We expect that the same anticrossing may be also present in CeCuAl$_3$, but more difficult to resolve as the first CEF excitation is rather low (1.3\,meV).\cite{D.T.Adroja2012}

\begin{figure}
\centering
\includegraphics[width=8.6cm]{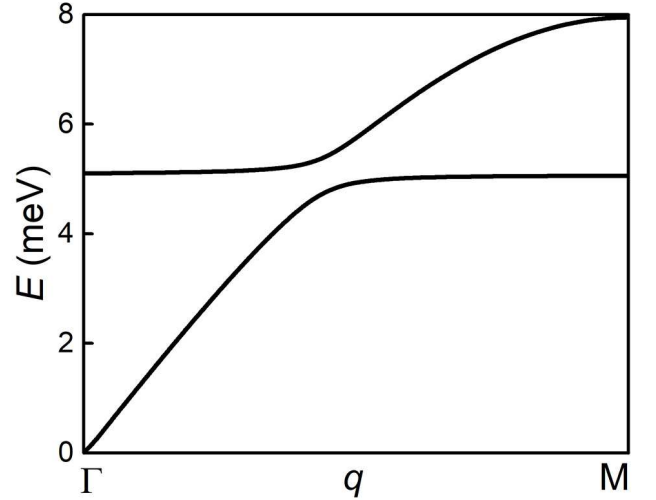}
\hspace{0.5cm} \caption{Dispersion curves of the coupled quadrupole-phonon excitations $\omega_{\mathbf{q}\pm}$ (see Eq.(14)) in the paramagnetic phase of CeAuAl$_3$ ($|\Gamma_6\rangle\rightarrow|\Gamma_7^1\rangle$), with $\Gamma=(0,0,0)$, M$=(0,0,1)$. The magneto-elastic coupling results in a clear anticrossing.}
\end{figure}

\section{IV. CONCLUSIONS}

In summary, we have studied the structural properties and lattice dynamics of the non-centrosymmetric heavy fermion compound CeAuAl$_3$ by first principles calculations. The irreducible representations for lattice modes at high symmetry points have been determined. Our results appear to contrast TOF neutron scattering experiments \cite{Adroja2015} which concluded that strong magnetoelastic interactions were absent in CeAuAl$_3$ as a bound state between CEF excitations and phonons could not be observed.

However, in our calculations we find that moderate magnetoelastic interactions are sufficient for the formation mixed-modes of phonons and quadrupole excitations. Using the theoretical deviation developed by Aksenov \textit{et al.} \cite{V.L.Aksenov1981}, a direct calculation of the matrix elements $V_{mn}(\mathbf{J},\mathbf{q})$ shows, that they are non-vanishing for $\Gamma_6^A-\Gamma_7^{1B}$ and $\Gamma_6^B-\Gamma_7^{1A}$, driving coupled quadrupole-phonon excitations even in the paramagnetic state. The analytic expression for the mixed excitation dispersions $\omega_{\mathbf{q}\pm}$ has been inferred from the poles of the phonon Green-function. In our recent inelastic neutron scattering experiments \cite{P.Cermak}, these mixed excitations have been observed directly, providing strong evidence for the importance of magnetoelastic interactions in CeAuAl$_3$.

\section*{ACKNOWLEDGMENTS}

This work was supported by the National Natural Science Foundation of China (No.11875238), and Science Challenge Project (No.TZ2016004). B.-Q. Liu acknowledges support by the China Scholarship Council. P. \v{C}erm\'{a}k, A. Schneidewind, and C. Pfleiderer acknowledge financial support under the joint DFG-GACR-project WI 3320/3-1, Czech Science Foundation Grant No. 17-04925J as well as DFG TRR80 (projects E1 and F2). We gratefully acknowledge helpful discussions with P. Thalmeier, M. Loewenhaupt, M. Wilde and P. Javorsk\'y, and computing time at the J{\"u}lich Supercomputer Centre on the general purpose system JURECA \cite{Jureca}.


\end{document}